\documentclass[usenatbib,useAMS]{mn2e}
\usepackage{url}
\usepackage{graphicx}

\usepackage{aas_macros}

\title[]{Do accretion discs regulate the rotation of young stars?}

\author[]{S.\,P.\ Littlefair$^{1}$, Tim Naylor$^{1}$, 
Ben Burningham$^{1}$, R.\,D.\ Jeffries$^{2}$\\
$^1$School of Physics, University of Exeter, Exeter EX4 4QL, UK\\
$^2$School of Chemistry and Physics, Keele University, Keele, Staffordshire, ST5 5BG, UK\\}

\date{\center{\Large Submitted for publication in the Monthly Notices of the
Royal Astronomical Society \\ 
\vspace{.5cm} \today}} 

\begin{document}
\maketitle

\begin{abstract} 
  We present a photometric study of $I$-band variability in the young
  cluster IC 348.  The main purpose of the study was to identify
  periodic stars. In all we find 50 periodic stars, of which 32 were
  previously unknown. For the first time in IC 348, we discover
  periods in significant numbers of lower-mass stars (M $<
  0.25$M$_{\odot}$) and classical T-Tauri stars. This increased
  sensitivity to periodicities is due to the enhanced depth and temporal
  density of our observations, compared with previous studies. The
  period distribution is at first glance similar to that seen in the
  Orion Nebula Cluster, with the higher-mass stars (M $>
  0.25$M$_{\odot}$) showing a bi-modal period distribution
  concentrated around periods of 2 and 8 days, and the lower-mass
  stars showing a uni-modal distribution, heavily biassed towards fast
  rotators. Closer inspection of the period distribution shows that
  the higher mass stars show a significant dearth of fast rotators,
  compared to the Orion Nebula Cluster, whilst the low mass stars are
  rotating significantly faster than those in Orion. We find no
  correlation between rotation period and $K-L$ colour or H$\alpha$
  equivalent width.
  
  We also present a discussion of our own IC 348 data in the context
  of previously published period distributions for the Orion Nebula
  Cluster, the Orion Flanking Fields and NGC 2264. We find that the
  previously claimed correlation between infrared excess and rotation
  period in the ONC might in fact result from a correlation between
  infrared excess and {\em mass}. We also find a marked difference in
  period distributions between NGC 2264 and IC 348, which presents a
  serious challenge to the disc locking paradigm, given the similarity
  in ages and disc fractions between the two clusters.

\end{abstract} 

\section{Introduction}
\label{sec:introduction}
The question of the early evolution of stellar angular momentum is an
interesting and essentially unsolved problem.  It is well known that a
substantial number of young stars show spin rates well below their
break-up speed \cite[e.g.][]{bouvier93}, despite the fact that these
stars should spin up as they contract towards the main sequence.
Furthermore, the angular momentum distributions of older stellar
clusters such as the Pleiades, and the simultaneous presence of
ultrafast rotators and slowly rotating stars amongst solar-type stars,
argue for a phase of angular momentum regulation amongst
pre-main-sequence stars \citep{barnes01,tinker02}. The most promising
mechanism to regulate the angular momentum of young stars is
disc-locking - a proposal first put forward by \cite{konigl91}. In the
disc locking theory, magnetic field lines connect the star to the
disc, enforcing synchronous rotation between the star and the material
in the disc at some radius, near where the magnetic field disrupts the
disc.

However, observational and theoretical support for disc locking is
still controversial. Whilst evidence exists that some young stars
possess the kilogauss-strength magnetic fields necessary to disrupt
the accretion flow \cite[e.g.][]{johns-krull99}, theoretical arguments
by \cite{safier98} and \cite{matt04} have called into question the
effectiveness of disc locking. Furthermore, \cite{hartmann02} argues
that, even if disc locking is effective, it may take almost 1 Myr to
affect the angular momentum of a young star.  Observationally, evidence
for disc locking is ambiguous. In the Orion Nebula Cluster (ONC), the
most heavily studied region to date, studies of rotation rates were
initially conflicting. Whilst \cite{herbst92} and later
\cite{herbst96b} found a bi-modal period distribution amongst the
young stars, \cite{stassun99} argued that the distribution was not
significantly different from uniform.  The source of this conflict has
now been resolved - the period distribution in the ONC is
mass-dependent. Whilst the brighter stars show a bi-modal
distribution, the fainter stars (which dominate the study of
\citealt{stassun99}) show a uni-modal distribution. 

In addition to a bi-modal period distribution, \cite{herbst02} also
report a strong correlation between rotation rate and infrared excess,
and suggest that the ONC therefore provides a strong argument for the
reality of disc locking.  This confirms the findings of earlier
studies, which found significant differences in rotation rates between
stars still surrounded by circumstellar discs and those that lack
evidence for discs \cite[e.g.][]{edwards93}.  However, other studies
of rotation rates fail to show statistically significant differences
between stars which show and stars which lack disc signatures
\cite[e.g.][]{stassun99,rebull01,rebull02,rebull04}. Furthermore, we
argue in this paper that the correlation between infrared excess and
rotation rate, as seen by \cite{herbst02}, might in fact be a
consequence of the dependence of rotation rate upon stellar mass
 (see section~\ref{subsec:p_v_disc}).  A lack
of any correlation between disc presence and rotation rate is a
serious challenge to disc locking.

Here we present a study of the photometric variability of stars in the
young cluster IC 348.  IC 348 is nearby and extremely young. Its
distance is 260 pc from {\em Hipparchos} parallaxes or 316 pc if
main-sequence fitting is used \citep{scholz99, herbig98}. Based upon
isochronal fitting of the pre-main sequence, the median age of the
cluster is between 1.3 and 3 Myr (depending upon the distance
adopted). The membership status of stars towards IC 348 is well
established by a major spectroscopic census \citep{luhman03}.  These
factors make IC 348 an ideal cluster in which to study the rotation of
young stars. Importantly, IC 348 is older than the well studied ONC
region, and therefore should allow a determination of timescales
involved in disc locking.  Indeed, IC 348 has already been
photometrically monitored, with the aim of determining rotation
periods \citep{herbst00,herbst04}. However, the relatively shallow
nature of these studies means that few periodic variables were
discovered, hampering the ability of those studies to make
statistically significant conclusions. The study presented here
re-visits IC 348 with deeper photometry with the aim of increasing the
number of known periodic variables in IC 348.

In section~\ref{sec:obs} we present the observations and data
reduction techniques applied.  Section~\ref{sec:var} characterises the
internal accuracy of our dataset, and describes some of the more
interesting irregular variables. Section~\ref{sec:per} describes the
techniques used to identify periodic variables and the results of our
study. Finally, in section~\ref{sec:discussion} we compare the period
distribution found for IC 348 with other published period
distributions, and draw our conclusions.

\section{Observations and Data Reduction}
\label{sec:obs}

\subsection{Observations}
\label{subsec:obs}
RGO $I$-band CCD images were taken with the CCD imager on the Jacobus
Kapteyn Telescope (JKT) on La Palma, equipped with a single
2048$\times$2048 pixel SITe array.  The imager covers approximately 14
$\times$ 14 arcmin on the sky, however only the central 9 $\times$ 9
arcmin is useable for accurate photometry, as the edges of the imager
suffer from strong vignetting. A single field, centred on IC 348
($\alpha = 3^h 44^m 28.5^s$, $\delta = +32^d 08^m 36^s$ J2000) was observed,
with data being taken on every night between 28/29th December 2002 and
9/10th January 2003, with additional imaging on the nights beginning
19th--22nd January 2003. The number of nights on which data were taken
totalled seventeen, and covered a baseline of twenty-six days. The
seeing varied between 0.8 and 3\arcsec, with a median and standard
deviation of 1.3\arcsec and 0.4\arcsec respectively.  Most nights were
affected by thin to heavy cirrus cloud, although the nights beginning
on 12th December 2002, 3rd January 2003 and 6th January 2003 were
photometric.  To increase the dynamic range of the resulting study, we
used exposure times of 2, 30 and 3$\times$300 seconds which were
repeated many times throughout a night. This yielded between 10 and 20
observations with the short and medium exposure times and between 30
and 60 observations with the long exposure time per night. In total
we obtained 227 useable short exposures, 230 useable medium exposures and
607 useable long exposures.

\subsection{Image Processing and Optimal Photometry}
\label{subsec:phot}

The individual frames were bias subtracted using a median stack of
several bias frames. In most cases bias frames taken on the same night
were used, with the exception of the 31st December and the 1st
January, where the bias frames were affected by stray light from the
dome.  The frames were then flat fielded using twilight sky flats
taken on one of the photometric nights. Only sky flats with peak
counts of less than 30,000 were used, to avoid non-linearity effects.
The CCD was affected by a number of bad columns and pixels. A bad
pixel mask was constructed for all frames by flagging all pixels which
deviated by more than 10$\sigma$ from the median of a ratio of two
flat fields. In addition, a smoothed version of a long exposure was
subtracted from the long exposure, and all pixels with a value lower
than 1000 were masked.  These procedures accounted for most bad
pixels, but left a number of bad columns unmasked. These columns were
identified by hand and added to the bad pixel mask.

The processed CCD frames were then analysed using optimal photometry,
as implemented by a slightly modified version of the {\sc cluster}
software described in detail by \cite{naylor02}. Advantages of this
approach over classical aperture photometry include better
signal-to-noise ratios, and robustly determined uncertainties for each
observation. Firstly, we summed the exposures from the night of 6th
January 2003 (the night which had the best seeing), after correcting
for positional shifts to form a deep $I$-band image which was then
searched for objects. Only stars with a signal-to-noise ratio
threshold of 15 in the summed image were retained, low enough to
detect all the stars for which good photometry would be obtained in
the 300 second exposures, but high enough to prevent spurious
detections. We note here that, as we are only interested in the high
signal-to-noise ratio data required to detect the small periodic
variations in T-Tauri stars, all stars with a mean signal-to-noise in
a single 300s image of less than 50 were removed from the final
catalogue.  This corresponds to a magnitude cut at I=18. The shifts
between each individual frame, and the summed $I$-band image were then
parameterised with a 6-coefficient solution, and optimal photometry
\citep[see][]{naylor98} was performed for every object in each frame.
Because the signal-to-noise of our images was high, we fitted the
positions of each object independently for each frame.

In classical aperture photometry, an aperture correction is used to
correct the flux measured in the small apertures used for the objects,
to the large apertures used for standard stars. There is an analogous
process in optimal photometry, known as ``profile correction''.
Bright, unsaturated stars in each frame were used to derive the
profile correction which was applied to the optimal photometry values.
Surprisingly, given the small field of view of the JKT imager, we
found that this profile correction must be allowed to vary as a
quadratic function of position, implying that the PSF of the CCD
imager on the JKT is spatially variable. The spatially varying
PSF of the CCD imager would affect relative aperture photometry too,
as the aperture correction would also need to vary with position.

An important aside arises regarding {\em relative} photometry. It is
often assumed that there is no need to account for this effect in
relative photometry, because the position of a star is usually
constant on the CCD. Although the aperture or profile correction may
indeed vary with position, the error so introduced therefore should
remain constant for any given object, and so the relative photometry
will still be correct. We find that this assumption is wrong because,
crucially, the seeing will vary between observations. At times where
the seeing is poor, the variable PSF of the instrument will be swamped
by bad seeing, and the aperture correction will be constant across the
frame. At times where the seeing is good however, the variable PSF of
the instrument becomes important again. If the aperture or profile
corrections are not allowed to vary as a function of position and time,
therefore, such behaviour will lead to implied variablility, even for
constant objects. It is therefore crucial to allow the aperture or
profile correction to be spatially variable, even for relative
photometry on small-field imagers like the JKT. Failure to account for this
affect likely explains the ``variable'' stars found around the edge of the
field in the recent study of \cite{lamm04}.

After profile correction, the photometric measurements were adjusted
for any difference in the airmass and transparency for each frame, by
determining a relative transparency correction from the bright stars.
A critical step in the determination of the transparency correction is
the identification of stars whose average magnitude will define the
reference brightness for each frame. Our reduction differs from that
described in \cite{naylor02}, in that this process was performed
iteratively. First, we used all stars with good measurements and
signal-to-noise ratios larger than 10 to define a reference
magnitude. Then we reject all stars with large scatter,
signal-to-noise ratios lower than 10 or a small number of good
measurements and recompute a reference magnitude. This process was
repeated until the $\chi^2_{\nu}$ of the remaining stars was less than
2. Before this process we added an additional, magnitude-independent
error of 0.01 to the results of each frame, in order to yield a plot
of $\chi^2_{\nu}$ versus signal-to-noise ratio that was flat and had a
modal value of approximately 1. The transparency correction was
applied to all observations, with airmasses ranging from 1 to 3. In
the absence of colour information, errors may be introduced into our
relative photometry by the colour dependence of extinction. The
following simulation was performed to estimate the importance of this
effect. By cross-correlation with the photometry of \cite{herbig98} we
find that the stars used to determine the transparency correction have
$R-I$ colours between 1.5 and 3, whereas the stars in our catalogue
have $R-I$ colours between 1 and 3.  We compared the $I$-band
magnitudes of two stars with $R-I$ values of 1.5 and 3.0.  Simulating
their spectra as linear fits between the fluxes at the mid-points of
the two bands and folding them through the responses of the Earth's
atmosphere, the telescope, filter and CCD, we found that their
relative $I$-band magnitudes changed by 3mmags between an airmass of 1
and 3. Hence we are confident this effect is negligible for the
purpose of our analysis.

An astrometric solution was achieved through comparison with a 2MASS
catalogue of the same region. A 6-coefficient solution to 144 stars
yielded a rms discrepancy in positions of 0.06 arcseconds. No
astrometric distortion model was included, and the position
discrepancies showed no systematic trends, implying that the JKT
imaging system is astrographic to an accuracy of at least 0.06
arcseconds.

There are three further departures from the methods outlined in
\cite{naylor02}, worthy of note.
\begin{enumerate}
\item{We have altered the way in which the $I$-band magnitude is
    ``flagged'' in order to describe the quality of a single data
    point. We now use character based flags according to the scheme
    laid out in \cite{bgb03}. In addition, we have added a flag to
    denote observations which fall within the heavily vignetted region
    of the CCD, which we label with the character R.}
\item{To reduce the effect of cosmic ray hits upon the lightcurves
    presented here, we remove such events by making an additional test
    for non-stellarity (following the methods outlined in
    \citealt{naylor02}) upon every star in each frame.  A star which
    is affected by cosmic ray hits will fail this non-stellarity
    test, and the measurement for this image is flagged with the character N.}
\item{The sky estimation technique is robust to the presence of nearby
    stars inside the sky box or smooth gradients in the sky
    background.  However, curvature in the sky background, as might be
    introduced near bright stars or in the vignetted regions of the
    CCD, will result in a histogram of pixel values that is highly
    assymetric and not well fit by the skewed Gaussian model we adopt.
    We therefore demand that the the $\chi^2_{\nu}$ of the fit to
    the sky histogram is less than 4, and that the skewness parameter
    is less than 0.3. If either of these conditions is not satisfied
    an ill-determined sky flag (I) is set.}
\end{enumerate}

\subsection{Transformation to a standard system}

\begin{figure}
\includegraphics[scale=0.37,angle=90,trim=50 0 0 80]{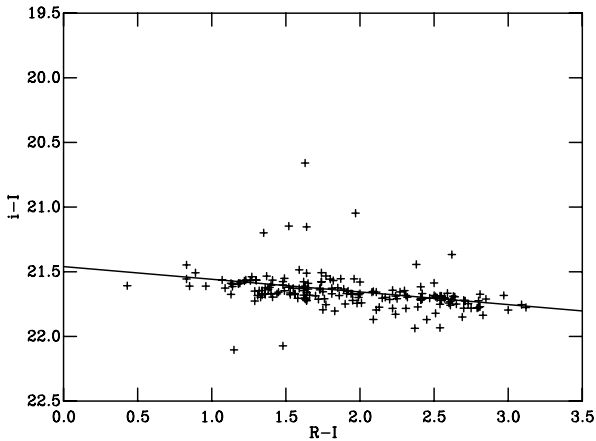}
\caption{Difference between our instrumental magnitude $i$ and standard 
magnitude on the Cousins system measured by \protect\cite{herbig98} as 
a function of colour. A linear least-squares fit to the data is shown.}
\label{fig:ivsi}
\end{figure}

Our aim in obtaining this dataset was to obtain high standard relative
photometry of IC 348. This meant that we did not obtain colour
information for each star at each epoch, and hence we are unable to tie
our photometry to a standard system directly. Instead, we tied our
photometry to the photometric system of \cite{herbig98}. The
\cite{herbig98} catalogue contains photometry of all the stars towards
IC 348, which was in turn tied to the Cousins system by observations
of standard stars from the list of \cite{landolt92}. The difference in
our instrumental magnitude $i$, against standard magnitude $I$, from
\cite{herbig98} is shown in figure~\ref{fig:ivsi}.
The linear, least-squares fit gives the following relationship between 
Cousins $I$-band magnitude and instrumental $i$:
\begin{equation}
I = i + 21.46 + 0.098 \times (R-I).
\end{equation} 
In the absence of colour information, we simply used an average $R-I$
value of 2.0 to convert our instrumental magnitudes to Cousins I.
Whilst not ideal, figure~\ref{fig:ivsi} shows that it provides
$I$-band magnitudes correct to around 0.3 mags. This should be borne
in mind when comparing this catalogue to others in the literature.

\subsection{Final Dataset}
\label{subsec:data}

The final data set consists of lightcurves for 169 stars, with I-band
magnitudes brighter than I=18.  Rather than combine data with very
different sizes of error bars, three lightcurves were produced for
each star, resulting from the 300, 30 and 2 second exposures. From the
300 second exposures, 138 stars have lightcurves with a median
signal-to-noise ratio of 10 or more, and at least 30 unflagged
datapoints. For the 30 second exposures, a further 15 stars satisfy
the same criteria. The 2 second exposure satisfy the same criteria for
an additional 9 stars.

The best quality lightcurve for each star, along with positions, mean
I-band magnitudes, and periodograms are available electronically via
Centre de Donn\'{e}es astronomiques de Strasbourg (CDS), and also from
the Cluster collaboration homepage
\url{(http://www.astro.ex.ac.uk/people/timn/Catalogues/description.html)}.

\section{Variability}
\label{sec:var}

\begin{figure}
  \includegraphics[scale=0.34,angle=-90]{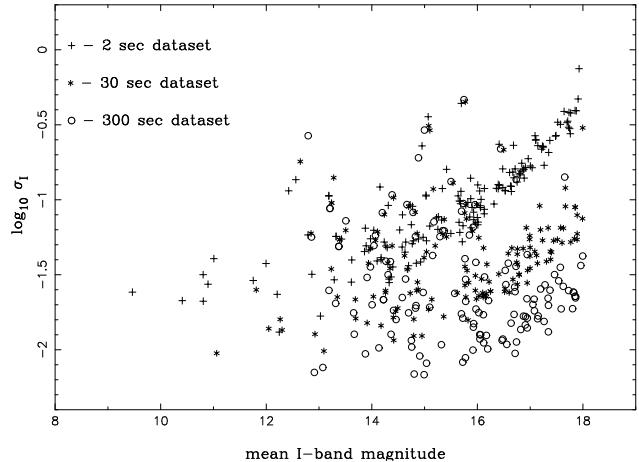}
\caption{Scatter in the photometry of stars as a function of brightness. Different
symbols correspond to data from different exposure times.}
\label{fig:rms}
\end{figure}
Figure~\ref{fig:rms} shows the RMS variability in magnitudes for our
stars, plotted as a function of magnitude. This plot shows the high
internal accuracy reached in our dataset. The 30 second and 300 second
exposures have an internal accuracy of around 0.01 mags. The 2 second
dataset is somewhat poorer, with an internal accuracy of only 0.02
mags. The most likely explanation for this is that the 2 second
dataset had relatively few stars suitable for use in determining the
transparency correction for each frame. Typically, around 17 stars
were available for this, but the number of suitable stars was
sometimes as low as five or six.

\begin{figure}
  \includegraphics[scale=0.37,angle=90,trim=50 0 0 80]{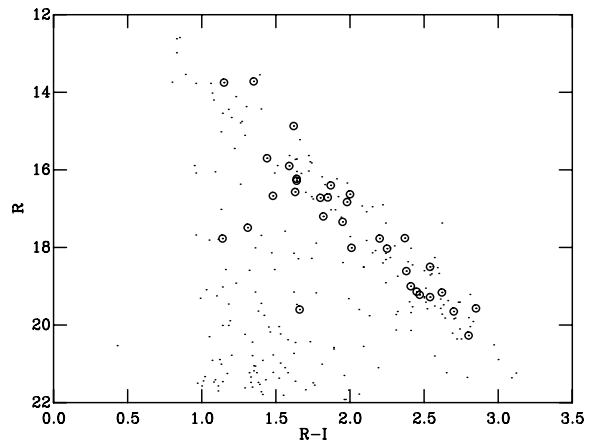}
\caption{Colour magnitude diagram of IC 348. Circles represent variable stars.}
\label{fig:varcmd}
\end{figure}
We could use the RMS variability as a means of detecting variable
stars, however the availibility of robust error determinations for our
measurements makes $\chi^2_{\nu}$ a more useful measure of
variability.  We have described a star as variable if $\chi^2_{\nu} >
2.5$ (the probability that this value of $\chi^2_{\nu}$ might arise by
chance in a non-variable star obviusly depends upon the number of good
data points, but is always less than 1\%). As an aside, we note that
variability is an excellent way of discovering pre-main sequence
stars. Figure~\ref{fig:varcmd} shows the $R$-band magnitude and $R-I$
colours (as determined by \citealt{herbig98}) of the variable stars in
our survey. The vast majority of the variable stars lie in the
pre-main sequence region.

\subsection{Irregular Variables}
\label{sec:irr}

Forty-three stars are irregular variable stars (showing $\chi^2_{\nu} >
2.5$ and with no detected period). The standard picture of variability
in T-Tauri stars (see \citealt{herbst94}, for example), predicts that
classical T-Tauri stars are more highly variable and erratic than the
weak-lined T-Tauri stars, because of the presence of variability from
accretion. As the magnitude distribution of CTTs and WTTs are similar
in our dataset, we can perform a direct comparison of variability
between the two groups to test this prediction.  We note here that we
have classified the stars as weak-lined or classical based upon the
H$\alpha$ equivalent width measurements presented in \cite{luhman03}.
Stars with H$\alpha$ equivalent widths greater than 10\AA\, are
classified as classical T-Tauri stars, otherwise a star is classed as
weak-lined.  

\begin{figure*}
  \includegraphics[scale=0.65,angle=-90]{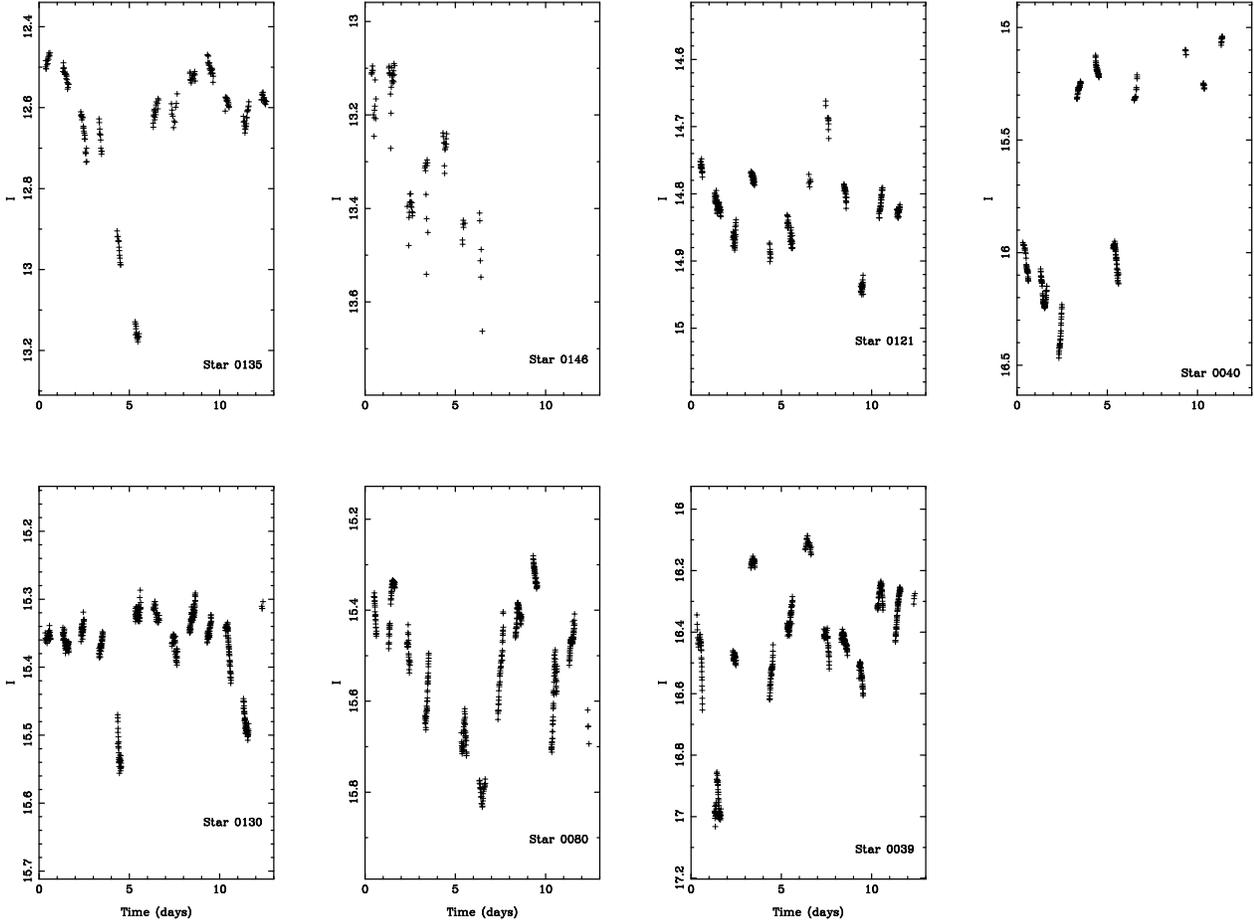}
\caption{Lightcurves of the most variable non-periodic CTTs. All classical T-Tauri stars
with $\chi^2_{\nu} > 20.0$ are shown.}
\label{fig:cttsvar}
\end{figure*}

Overall, the variability seems to support the standard picture of
variability in T-Tauri stars.  The classical T-Tauri stars are more
highly variable than the weak-lined T-Tauri stars, with the CTTs
showing an average standard deviation of 0.1 mags, compared with 0.05
mags for the WTTs.  However, the CTTs are just as likely to be
non-variable as the WTTs - the ratio of CTTs to WTTs amongst the
non-variable stars is $0.41\pm0.1$, which is not different from the
ratio of CTTs to WTTs in the cluster as a whole.
Figure~\ref{fig:cttsvar} shows lightcurves for the most highly
variable classical T-tauri stars.  The variability reaches amplitudes
as high as 1.5 mags (star 40), and is characterised by erratic
flickering. The most likely source of this variability is irregular
accretion onto the young star.

\begin{figure*}
  \includegraphics[scale=0.65,angle=-90]{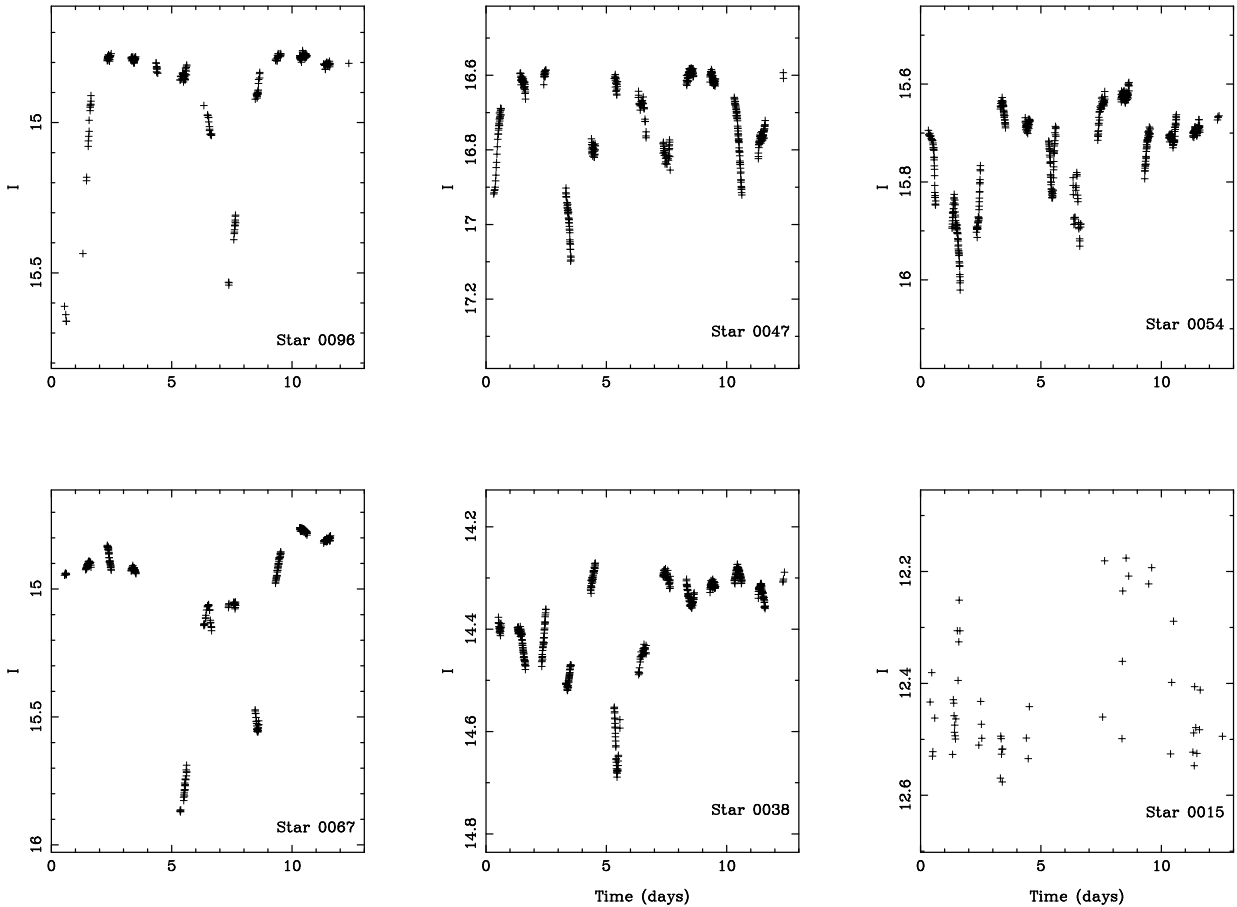}
\caption{Lightcurves of the most variable non-periodic WTTs. All weak-lined T-Tauri stars
with $\chi^2_{\nu} > 20.0$ are shown.}
\label{fig:wttsvar}
\end{figure*}

Somewhat surprisingly then, large amplitude variability is not
restricted to the classical T-Tauri stars. Out of the 43 irregular
variables found, 23 are weak T-Tauri stars, and six weak-lined T-Tauri
stars show large amplitude variability ($\chi^2_{\nu} > 20.0$).  These
stars are shown in figure~\ref{fig:wttsvar}. The amplitudes and
characteristics of the variability in these stars are similar to those
seen in the CTTs. Two stars are unusual and worthy of further mention.
Star 67 (Herbst 47) was seen by \cite{herbst04} to change from being
periodic with a period of 8.4 days to showing large erratic dips of up
to 1 mag.  These dips are interspersed with long intervals near
maximum brightness. Our lightcurve shows the same erratic behaviour.
This behaviour is characteristic of UX Ori stars (see
\citealt{herbst94}).  UX Ori type behaviour is usually attributed to
occultations of the central star by disc material, and it is therefore
highly unusual for such behaviour to be seen in a weak-lined T-Tauri
star. Star 96 (Herbst 73) was included as a periodic star in
\cite{herbst04}, with a period of 7.6 days. The star is not included
in our periodic sample, because the $\chi^2_{\nu}$ with respect to a
sinusoidal fit is too large. However, given that we find a period of
7.2 days for this star, it clearly is periodic, and furthermore the
period is stable over several years. It is difficult to explain the
behaviour of star 96 in terms of starspots and stellar rotation,
however.  The star remains at maximum brightness for over four days,
then suddenly drops in flux by over 50\%, recovering its flux around 3
days later. We believe the behaviour of this star may be due to
eclipses by extended, long-lived structure in an accretion disc around
the star. Similar behaviour (albeit with a much longer period) has
been explained in this manner for the star KH 15D \citep{hamilton01}.

It is tempting to link the variability of stars 96 and 67 to the
presence of a circumstellar disc. The presence of a circumstellar disc
in these objects is confirmed by the presence of a $K-L$ excess
\citep{haisch01}. However, these are weak-lined T-Tauri stars, which
are traditionally thought to lack discs. In fact, a strong body of
evidence is emerging that a significant fraction of stars classified
as weak-lined T-Tauri stars do possess discs: the eclipses in KH
15D are strongly indicative of eclipses by disc material
\citep{hamilton01} and infrared excesses have been seen around a large
number of weak-line T-Tauri stars \citep{hetem02}. In addition,
\cite{littlefair04} claim that the composite spectrum effect, first
seen by \cite{ba92} in a group of four weak-lined T-Tauri stars, is
caused by high accretion rates, which necessarily implies the presence
of an accretion disc.  


\section{Periodic Variables}
\label{sec:per}

\subsection{Detection of Periodic Variables}
\label{subsec:detect}

In identifying periodic variables, our aim was to detect the largest
number of periodic variables possible, whilst keeping the fraction of
spurious detections suitably low. Traditional methods of estimating
the significance of a detected signal \citep{horne86b,nemec85} are not
applicable to our dataset. The method of \cite{horne86b} is strictly
applicable only to evenly spaced, uncorrelated datapoints
\citep[see][]{herbst96}, whilst the method of \cite{nemec85} is not
valid for correlated datapoints. Night-to-night variations in our data
(e.\,g.\,transparency corrections, non-periodic stellar variablility)
will make our datapoints strongly correlated; our datapoints are, by
necessity, unevenly spaced.

As there is no formally correct way to calculate the significance of
our detected periods, we use a modified version of the procedure
detailed in \cite{herbst00}. Two periodograms are calculated for each
star, using the method formulated by \cite{horne86b}. One periodogram
is based upon the whole lightcurve, the other based upon a modified
dataset consisting of the weighted mean of each night's data. This
latter dataset is much closer to the ideal of evenly spaced,
uncorrelated data, and it is reasonable to use it with the method of
\cite{horne86b} to calculate a false alarm probability (FAP) for the
period found in the original dataset. In principle it is now possible
to differentiate between real and spurious periods by making a cut at
some value of the FAP. What value of the FAP to use is necessarily a
subjective decision. Naylor and Littlefair independently classified
the period detections based upon visual inspection of the folded
lightcurves from the 300s dataset. The period detections were grouped
into three categories; certain, uncertain and spurious. The
distribution of the FAP for each category was examined. In both cases,
few ``certain'' detections had FAPs greater than 0.08. However, there
were a substantial number of ``spurious'' and ``uncertain'' detections
below this point.

We were reluctant to use a lower FAP as our cut-off, as this meant
throwing away many detections independently classified as ``certain''.
Instead, we looked at the lightcurves of the ``spurious'' and
``uncertain'' stars which had passed our FAP cut-off to see if there
were some other criteria we could use to reject these stars. The
majority of ``spurious'' and ``uncertain'' detections were lightcurves
in which the amplitude of any variability was small or negligible. A
small number of ``spurious'' and ``uncertain'' detections were highly
erratic, large amplitude variables, and a further number were stars
whose detected periods were too long to be classed as ``certain''.
Additionally, some stars had too few data points, or the folded
lightcurves covered only a small fraction of the rotational cycle. To
filter out these objects we computed a third periodogram for all
stars, using a sinusoid fitting technique. At each period to be
considered, a sine-curve is fit to the data, and the resulting value
of $\chi^2_{\nu}$ is calculated. The detected period is taken to be
the period which minimises the value of $\chi^2_{\nu}$. After
calculating this periodogram, we rejected those stars with
$\chi^2_{\nu} > 15$\footnote{To ensure that the $\chi^2_{\nu}$ cut-off
  was not rejecting stars which were clearly periodic but
  non-sinusoidal, all the stars rejected because of $\chi^2_{\nu}$
  criteria were inspected by hand. Only one periodic object, star 96,
  was rejected on the basis of $\chi^2_{\nu}$. This star is discussed
  in detail in section~\ref{sec:irr}} , stars with periods longer than
14 days, stars whose lightcurves had less than 50 data points, or
whose folded lightcurve had gaps of more than 0.25 of a rotational
cycle. We also characterised the amplitude of variability by the
amplitude of the sine-fit $a$. Stars with $a \ge 18\sigma_m/
\sqrt{N_p}$, where $N_p$ is the number of data points, and $\sigma_m$
the median uncertainty in flux, were rejected. Also rejected were stars with
periods between 0.9 and 1.1 days.

The above procedure, derived from the 300s lightcurves alone, was
applied, without modification, to all three datasets.  The 300 second
lightcurves yielded 42 periodic variables. The 30 second lightcurves
yielded 37 periodic variables, of which 8 were not found in the 300
second dataset. These 8 stars had lightcurves with a large number of
saturated points in the 300 second dataset. The 2 second lightcurves
yielded 18 periods, all of which were present in either the 300 second
or 30 second dataset.  As a check that our period detection process
was not yielding significant numbers of spurious detections, we
examined the periods of those stars which shared detections across
datasets. In all cases, the periods agreed to within 10\% of each
other. For the purposes of this paper, the period adopted was the
value computed from the longer exposure dataset, via the sine-fitting
method.  In total, 50 periodic variables were identified. These stars
are shown in table~\ref{tab:pers}.  Periodograms from the
sine-fitting method are available online, as are the folded
lightcurves.

\begin{table}
\caption[]{Periodic variables in IC 348. $\chi^2_{\nu}$ is determined with respect to the best fitting sinusoid. H$\alpha$
equivalent width measurements are taken from \protect\cite{luhman03}. A value of 0.0 indicates that no H$\alpha$ emission
was detected or measurable. The masses are determined using the method described in 
section~\protect\ref{subsec:us_v_herbst}.}
\begin{center}
\begin{tabular*}{3.5in}{@{\extracolsep{-0.1cm}}rrrrrcrr}
& & & & & & &\\
\hline
\multicolumn{1}{c}{Star} & \multicolumn{1}{c}{Cohen}& \multicolumn{1}{c}{Period} & \multicolumn{1}{c}{$I$} 
& \multicolumn{1}{c}{$\chi^2_{\nu}$}& \multicolumn{1}{c}{SpT} & EW H$\alpha$ & Mass \\
\multicolumn{1}{c}{ID}& \multicolumn{1}{c}{ID}& \multicolumn{1}{c}{(days)} & & & & 
\multicolumn{1}{c}{(\AA)} & ($M_{\odot}$) \\
& & & & & & &\\
\hline
\hline
& & & & & & &\\

  4 & 50 &  5.36 & 13.338 & 0.52 & K5    &  1.5 &  \\
  5 & 88 & 13.69 & 15.747 & 5.47 & M2    & 22.0 &  0.33\\
  7 & 69 &  3.27 & 15.329 & 4.73 & M4.5  &  6.5 &  0.17\\
 10 & 37 & 14.00 & 14.332 & 6.31 & M4    &  5.5 &  0.18\\
 11 & 34 &  8.43 & 14.615 & 4.97 & M1    & 90.0 &  0.43\\
 20 & 86 &  1.39 & 15.902 & 8.26 & M5.25 &  3.0 &  0.16\\
 21 & 133&  6.31 & 15.076 & 1.51 & M0    & 51.0 &  0.53\\
 25 &    &  1.33 & 17.324 & 0.97 & M4.75 &  3.0 &  0.19\\
 31 & 39 & 10.16 & 14.199 & 0.69 & M2.5  &  4.0 &  0.28\\
 35 &    &  1.73 & 17.827 & 1.08 & M5.75 &140.0 &  0.13\\
 36 &    &  1.50 & 17.712 & 1.11 & M5.25 &  7.0 &  0.16\\
 41 & 82 &  3.74 & 14.732 & 0.45 & M4.75 &  4.0 &  0.17\\
 45 & 16 &  5.21 & 13.189 & 1.58 & K6    &  3.9 &  0.44\\
 48 & 60 &  6.96 & 14.643 & 1.97 & M1.5  &  0.0 &  0.38\\
 50 & 96 &  1.52 & 16.095 & 0.45 & M5    &  5.0 &  0.17\\
 52 & 98 &  3.58 & 16.935 & 6.03 & M4.25 & 50.0 &  0.18\\
 55 & 26 &  3.01 & 12.801 & 1.15 & K0    &  0.0 &  1.34\\
 56 & 81 &  3.98 & 14.768 & 1.15 & M2    &  8.0 &  0.32\\
 58 & 105&  1.73 & 15.703 & 0.75 & M5    &  7.0 &  0.17\\
 60 & 14 &  1.63 & 12.203 & 0.70 & G8    &  0.0 &  \\
 65 &    &  0.57 & 17.550 & 2.18 & M7.5  & 15.0 &  0.06\\
 69 & 46 &  2.09 & 14.389 & 0.54 & M3.25 &  3.0 &  0.23\\
 70 & 43 &  6.68 & 14.344 & 0.51 & M3    &  5.0 &  0.24\\
 71 &    &  1.77 & 17.245 & 0.72 & M5.75 &  6.0 &  0.15\\
 72 & 61 &  2.18 & 14.815 & 1.46 & M2    & 47.0 &  0.33\\
 73 & 45 &  2.39 & 13.883 & 0.40 & K3    &  0.0 &  1.01\\
 74 & 12 & 10.81 & 12.241 & 0.76 & K2    &  0.0 &  1.27\\
 77 & 27 &  2.65 & 11.755 & 0.86 & G6    &  0.0 &  \\
 78 & 108&  2.52 & 15.689 &11.45 & M4.75 & 30.0 &  0.16\\
 82 & 29 &  7.93 & 14.049 &12.14 & K7    & 68.0 &  0.40\\
 83 & 142&  1.68 & 16.176 & 0.55 & M5.25 &  0.0 &  0.16\\
 84 & 70 & 13.04 & 15.076 & 1.82 & M1.25 & 10.5 &  0.15\\
 85 & 30 &  7.55 & 14.184 & 1.09 & M4.75 &  0.0 &  0.31\\
 97 & 74 &  2.25 & 16.032 & 0.46 & K6.5  &  4.5 &  0.17\\
 98 & 41 &  7.12 & 13.460 & 0.52 & M4.75 &  1.4 &  0.57\\
100 & 122&  1.54 & 16.676 & 0.85 & M3.5  &  4.5 &  0.19\\
106 & 134&  8.56 & 15.836 & 1.47 & K5    &  9.0 &  0.21\\
111 & 51 & 12.43 & 13.970 & 0.60 & M1    &  1.5 &  0.51\\
115 & 135&  7.48 & 15.933 & 8.23 & M2.25 &  3.0 &  0.27\\
117 & 52 & 11.22 & 14.244 & 1.00 & M0    &  2.0 &  0.52\\
126 &    &  1.58 & 16.890 & 0.71 & M5    &  7.0 &  0.18\\
127 & 64 &  6.84 & 15.134 & 0.93 & M2.5  &  8.0 &  0.28\\
131 & 21 &  9.11 & 13.624 & 0.71 & M1    &  3.5 &  0.38\\
141 & 31 &  3.08 & 14.021 & 1.23 & M4.75 & 10.0 &  \\
143 & 127&  2.19 & 15.756 & 3.84 & M5.75 &110.0 &  0.14\\
144 & 32 &  4.90 & 15.275 & 1.39 & M1    &  5.0 &  0.39\\
147 &    &  1.55 & 17.644 & 3.86 & M5.5  &200.0 &  0.16\\
178 & 107&  1.90 & 16.028 & 0.57 & M5.25 & 74.0 &  0.15\\
179 &    &  1.79 & 17.265 & 1.02 & M4.75 &  5.0 &  0.19\\
188 & 137&  3.40 & 16.927 & 0.60 & M7.25 & 45.0 &  0.09\\
& & & & & & &\\
\hline
& & & & & & &\\
\end{tabular*}
\end{center}
\label{tab:pers}
\end{table}

\subsection{Comparison with previous work}
\label{subsec:us_v_herbst}

\begin{figure}
  \includegraphics[scale=0.35,angle=-90]{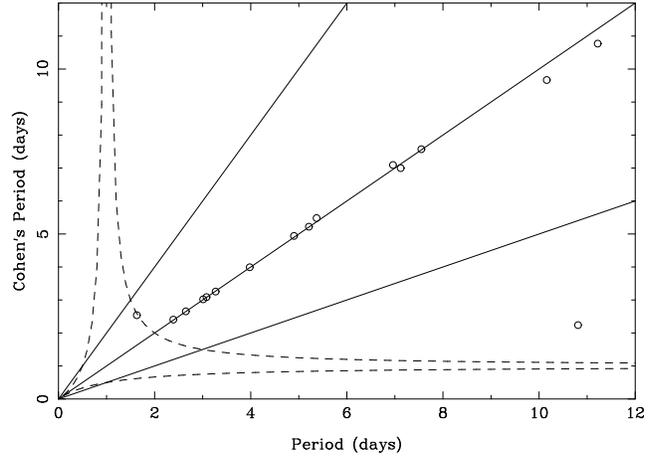}
\caption{A comparison between period determinations in this study and the study of 
  \protect\cite{herbst04}. The straight lines represent a 1:1
  agreement, and factors of two disagreement between the period
  determinations. A factor of two disagreement may arise if the
  harmonic of a period is detected instead of the actual period. The
  dashed lines show periodicities which might be expected due to
  beating between a real period and the 1 day sampling interval.}
\label{fig:uvh}
\end{figure}

Prior to this work, data on rotation periods in IC 348 have been
published in \cite{herbst04}. This study is based on observations
spread over five observing seasons with the 0.6m telescope at the Van
Vleck Observatory, and yields total of 28 known periodic stars (a
subset of this dataset was published in \cite{herbst00}). Of the 50
periodic stars presented here, 18 were previously known, and the
remaining 32 are new. Our study therefore more than doubles the number
of known periodic variables in IC 348.  Ten periodic stars from
\cite{herbst04} were not recovered in this study, seven of which are
within our field of view. Five of these seven stars were not
identified as periodic because they fail our false alarm probability
criterion.  Of these five stars, four stars show periods which either
agree with \cite{herbst04}, or are explained by aliasing or doubling
(see figure~\ref{fig:uvh}). This fact suggests that our selection
criteria for periodic stars is more conservative than that employed by
\cite{herbst04}\footnote{This conclusion is not affected by the fact
  that we have discovered many additional periods to those found by
  \cite{herbst04}. The new periods presented here are predominantly in
  low mass stars, for which the photometry of \cite{herbst04} would
  not have been sufficiently deep to detect periods.}  The remaining
two stars (stars 96 and 67) were rejected because they failed our
$\chi^2_{\nu}$ criterion. These stars are discussed in detail in
section~\ref{sec:var}.

The periods found for the 18 stars in common between this study and
the study of \cite{herbst04} are shown in figure~\ref{fig:uvh}. The
large majority of period determinations agree. This suggests that
rotation periods in young stars are stable over periods of five years
or so, confirming the results seen by \cite{herbst04}.
Figure~\ref{fig:uvh} also allows us to estimate the uncertainties on
our period determinations. Below periods of 10 days, the period
determinations agree to better than 0.1 days. We conclude that our
periods are at least this accurate. The plot shows increasing scatter
with increasing period. This is due to the comparatively short
monitoring time in this study, leading to a lower accuracy in periods
longer than 10 days. One period is consistent with a beat period
between the real period and the sampling interval of $\sim$1 day, and
a single star shows a disagreement in the claimed periods. This star
(star no.\,74 in this paper; Herbst's star no.\,12) exhibits a period
of 2.2 days in \cite{herbst04}, and 10.8 days in this study. Whilst we
are unable to explain the difference in periods for this single star,
the agreement in periods between the majority of stars gives us
confidence in our period determination procedures.

We were able to allocate masses using the spectroscopy of
\cite{luhman03} and the isochrones of \cite{dam97} for all 32 of the
newly-found periodic objects. The vast majority of these objects are
low mass stars. In fact, 87\% are lower than 0.25 M$_{\odot}$.  This
contrasts strongly with the study of \cite{herbst04}, where only a
single object is below 0.25 M$_{\odot}$.  The increased depth of our
study is a simple result of the larger aperture telescope used (1-m
versus 0.6-m), and the site quality of La Palma.  Also remarkable is
the fraction of classical T-Tauri stars discovered to be periodic in
our survey. \cite{herbst04} notes that in their photometric study,
none of the periodic stars found are classical T-Tauri stars, and
suggest that this might strongly bias the results of the survey. In
contrast, we have found periods in 15 classical T-Tauri stars. In fact
the ratio of classical to weak-lined T-Tauri stars amongst our
periodic sample is $0.43 \pm 0.13$, which is not significantly
different from the ratio of classic to weak stars in IC 348 as a
whole. It is apparent that our photometric study has been markedly
more succesful in detecting periods in classical T-Tauri stars than
the study of \cite{herbst04}.  Over half of the classical T-Tauri
stars in which we find periods are sufficiently bright that
\cite{herbst04} should also have found periods for these stars. The
most likely explanation for this difference is that our study has a
very high temporal density of observations, with typical observation
spacings of less than ten minutes, and continuous coverage for several
nights.  The study of \cite{herbst04} has much lower temporal density,
and several nights can elapse between observations. Because Classical
T-Tauri stars are erratically variable on timescales of a few nights,
this variability masks the periodic signal in the study of
\cite{herbst04}.  The dense spacing of our observations means that
this effect is less marked for our survey.  We conclude that
photometric studies for rotation rates in young stars must have a high
temporal density if they are not to be biased by a lack of Classical T
Tauri stars.

\subsection{Period distribution}
\label{subsec:perdist}
\begin{figure}
  \includegraphics[scale=0.37,angle=90,trim=50 0 0 80]{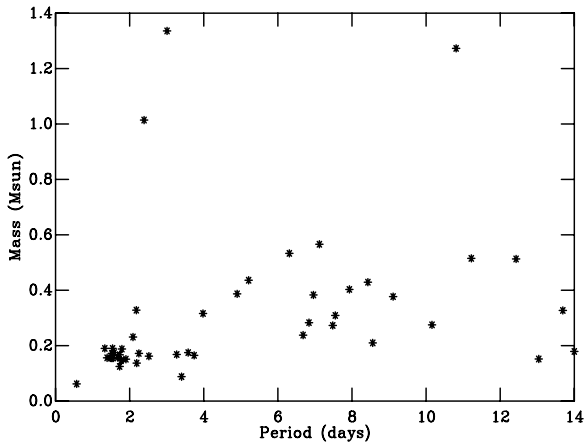}
\caption{Stellar mass against rotation period.}
\label{fig:mass_v_per}
\end{figure}

Figure~\ref{fig:perall} shows the period distribution of stars in our
sample, where we have split the stars into two groups; those with
masses above 0.25 M$_{\odot}$ and those below this mass.  For stars
with masses greater than $0.25 M_{\odot}$, the period distribution in
IC 348 appears bi-modal, with peaks around 8 days and 2 days although
applying the Hartigan dip test \citep{hartigan95a,hartigan95b} to the
distribution shows that the bi-modality is not statistically
significant. The period distribution for the high mass stars resembles
the period distribution seen for high-mass stars in the ONC.  A K-S
test on the periods of the high mass stars in IC 348 and the ONC gives
only a 55\% probability that the samples were not drawn from the same
parent distribution. Both clusters also show broadly similar behaviour
for stars with masses below 0.25 $M_{\odot}$ (although differences
exist, see below for details). For these stars, the period
distribution is clearly uni-modal, and the average rotation rate is
much higher, with most stars having periods around 1--2 days. A strong
correlation between stellar mass and rotation rate is also evident our
study of IC 348 (figure~\ref{fig:mass_v_per}). A Spearman Rank
Correlation test between stellar mass and rotation period rejects a
hypothesis of independence with a confidence of 99.996\%. This result
shows that the strong mass dependence of rotation rate seen in the ONC
\cite{herbst02} may well be a common feature of young stellar
populations. Depsite the broadly similar behaviour of rotation period
with stellar mass, there is evidence that the rotation period
distributions of the low-mass stars in IC 348 and the ONC are
different. The low-mass stars in the ONC have a tail of slower
rotators extending to periods of over 10 days, whereas the low-mass
stars in IC 348 are much more strongly grouped at around 1--2 days.  A
K-S test of the low mass period distributions in IC 348 and the ONC
finds a 98\% probability that they are drawn from different parent
distributions.  In section~\ref{sec:discussion} we present a full
discussion of these results, and also compare the rotation period
distribution in IC 348 with those in NGC 2264 and the Orion flanking
fields, as well as the ONC.

\subsection{Rotation Period and Disc Indicators}
\label{subsec:p_v_disc}

The extensive study of rotation rates in the ONC performed by
\cite{herbst02} showed a strong correlation between rotation period
and infrared excess, as judged by $I-K$ excess. Such a strong
correlation would be strongly suggestive that the rotation period
distribution in the ONC can be explained by disc locking. However, it
may well be that this is merely a secondary correlation, which arises
because of the strong dependence of $I-K$ excess with {\em stellar
  mass}. The level of $I-K$ excess depends upon a number of factors;
disc mass, inclination angle, inner disc hole size, and disc
structure.  It also depends strongly upon stellar mass, with excesses
around high mass stars being much stronger than excesses around low
mass stars \citep{h98}. What this means for the ONC is that the low
mass stars, which are rotating rapidly, will necessarily exhibit
smaller $I-K$ excesses than the high mass stars, which rotate more
slowly. Such an effect could easily be responsible for the apparent
correlation between rotation rate and infrared excess. The claims that
the ONC offers strong support for disc locking should therefore be
interpreted with caution.

\begin{figure}
  \includegraphics[scale=0.37,angle=90,trim=50 0 0 80]{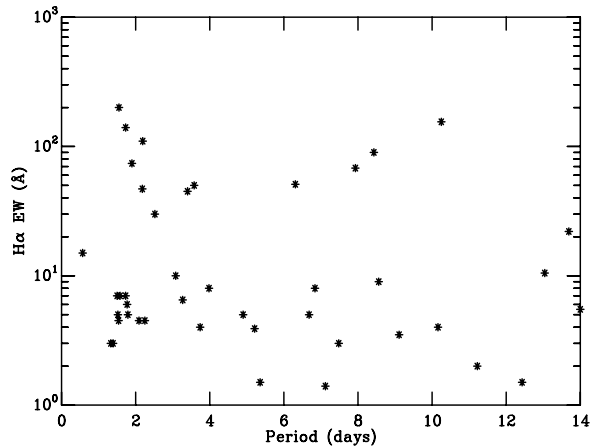}
\caption{H$\alpha$ equivalent width against rotation period. Only 
those 43 stars with a non-zero H$\alpha$ equivalent width in 
\protect\cite{luhman03} are shown here.}
\label{fig:hal}
\end{figure}

As we find a similar period distribution in IC 348 to that seen in the
ONC, it is pertinent to ask if we also see a correlation between
rotation period and signs of disc presence or accretion.
A plot of rotation period against H$\alpha$ equivalent width is shown
in figure~\ref{fig:hal}.
\begin{figure}
  \includegraphics[scale=0.37,angle=90,trim=50 0 0
  80]{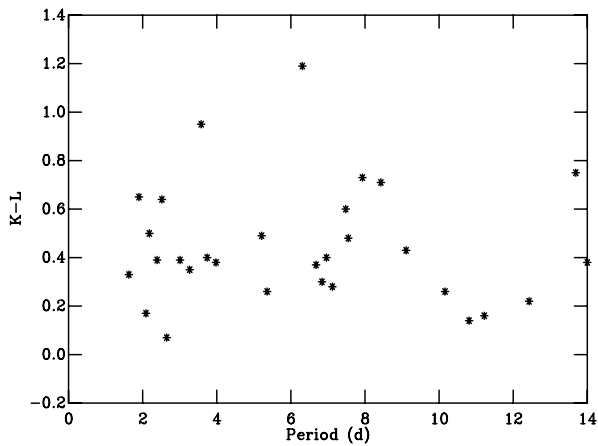}
\caption{$K-L$ colour against rotation period. $K-L$ colours are taken
from \protect\cite{haisch01}.}
\label{fig:KL}
\end{figure}
Strong H$\alpha$ emission is a commonly used sign of ongoing accretion
onto the central star.  We might therefore expect a correlation
between H$\alpha$ equivalent width, and the rotation period of the
star, in the sense that actively accreting stars are more likely to be
locked to their discs and hence be slow rotators. No such correlation
is seen.  In fact, a Spearman Rank Correlation test on the two
variables rejects the null hypothesis of independence with a
probability of only 30\%.  This result is confirmed by a K-S test of
the period distributions of the classic and weak T-Tauri stars, which
gives only a 3\% chance the two were drawn from different parent
distributions.  A similar result arises if we consider the dependence
of rotation period against infrared excess, as measured by
\cite{haisch01} using a $JHKL$ colour-colour diagram.  A star is
defined as a $K-L$ excess source if its $K-L$ colour places it to the
right of the reddening locus of a star at the end of the stellar main
sequence \citep[see][for example]{haisch01}. A K-S test of the period
distributions of $K-L$ excess sources and sources with no excess gives
a 7\% chance that they are drawn from different parent
distributions. Likewise, a Spearman Rank Correlation of rotation
period against $K-L$ colour gives a 27\% chance that the variables are
dependent (see figure~\ref{fig:KL}). We therefore do not find a strong
correlation between indicators of inner disc presence and rotation
period, nor do we find a strong correlation between rotation period
and accretion, as measured by H$\alpha$ equivalent width.

Smaller sample size is not sufficient to explain the absence of any
correlation.  To see if a correlation of the strength found in the ONC
by \cite{herbst02} should be detectable in our sample we randomly
selected 50 stars from their dataset, and computed the Spearman Rank
Coefficient of $\Delta (I-K)$ against rotation period. This was
repeated many times, and in only 1\% of tests was the absolute value
of the Spearman Rank Coefficient smaller than that measured between
$K-L$ colour and rotation period for IC 348. We conclude that our
sample size is large enough to detect a correlation between $K-L$
colour and rotation period, provided that correlation was at least as
strong as that seen by \cite{herbst02} between $\Delta (I-K)$ and
rotation period. In fact, we might expect that the correlation should
be even stronger when using $K-L$ as a diagnostic of disc presence, as
we might expect $K-L$ to be more sensitive to the presence of discs.
The most likely explanation as to why we see no correlation is that
the $K-L$ {\em colour} is not mass dependent in the same way as
$\Delta (I-K)$. This is because lower mass stars will exhibit smaller
$K-L$ excesses, but also larger intrinsic $K-L$ colour, and these two
effects will cancel each other out to some extent.

\section{Discussion}
\label{sec:discussion}
\begin{figure*}
  \includegraphics[scale=0.7]{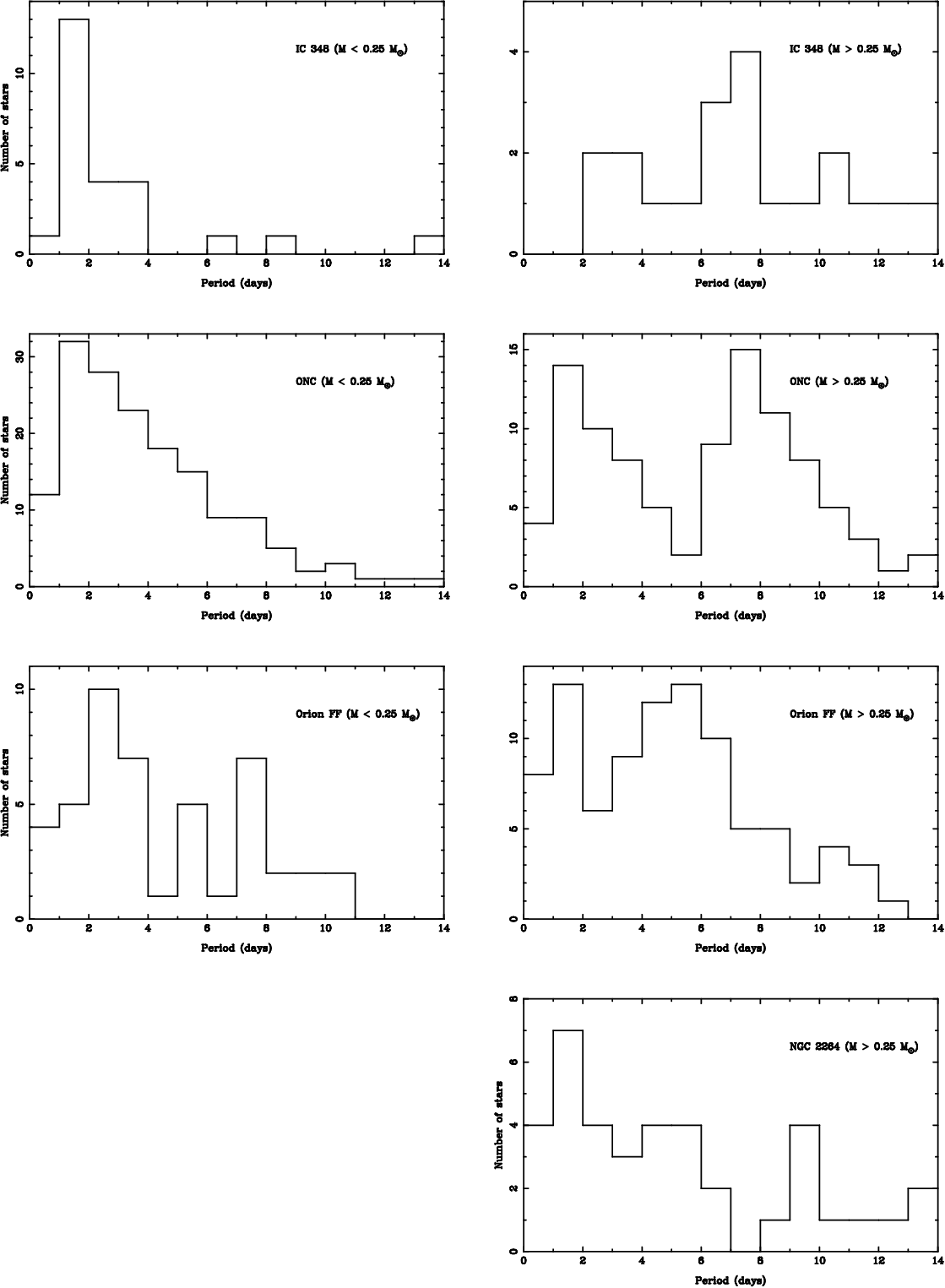}
\caption{Distribution of periods in IC 348 compared with the distributions 
  for the ONC (\protect\citealt{herbst02}), Orion flanking fields
  (\protect\citealt{rebull01}) and NGC 2264
  (\protect\citealt{rebull04}). Masses in the ONC have been assigned
  by \protect\cite{hillenbrand97} using the isochrones of
  \protect\cite{dam94}.  Masses in IC 348 have been assigned using the
  effective temperatures and luminosities of \protect\cite{luhman03},
  and the isochrones of \protect\cite{dam97}.  Identical period
  distributions are produced for IC 348 if the \protect\cite{dam94}
  isochrones are used. Masses for the stars in NGC 2264 were assigned
  by \protect\cite{rebull04}, following the method of
  \protect\cite{hillenbrand97}.  We assigned masses to the stars in
  the Orion Flanking fields using the spectral types and extinctions
  from \protect\cite{rebull01}, and following the method of
  \protect\cite{hillenbrand97}.  }
\label{fig:perall}
\end{figure*}

\subsection{Stars with masses greater than 0.25 M$_{\odot}$}
\label{subsec:himass}

The rotation period distributions for IC 348 (this sample), the ONC,
the Orion flanking fields (OFF) and NGC 2264 are shown in
figure~\ref{fig:perall}.  We first discuss the period distributions of
the higher mass ($M > 0.25M_{\odot}$) stars.  Above 0.25M$_{\odot}$,
the ONC period distribution is bi-modal, with peaks at 2 days and 8
days. This bi-modality is highly significant.  The Hartigan dip test
\citep{hartigan95a, hartigan95b}, gives only a 0.4\% chance that the
ONC distribution is consistent with a unimodal distribution.  This
bi-modal distribution, and the correlation between infrared excess and
rotation rate, led \cite{herbst02} to conclude that disc locking
explained the period distribution in the ONC.  By modelling the period
distribution in the ONC, and assuming that disc locking held stars at
periods near to eight days, \cite{herbst02} showed that the period
distribution in the ONC was consistent with 20--50\% of stars in the
ONC being locked to their discs. \cite{herbst02} argue that this is
because stars are released from disc locking after around 1 Myr.
However, the period distribution is also consistent with stars {\em
becoming} locked to their discs after around 1 Myr. In the discussion
to follow, we will refer to these as scenarios one and two,
respectively. There may be some justification for preferring scenario
two, as \cite{hartmann02} suggests that disc locking can only
spin-down stars as rapidly as angular momentum can be removed from the
inner disc, implying that young stars may indeed not be locked to
their discs for the early stages of their lives. However, the claims
of \cite{hartmann02} are challenged by \cite{matt04a}, who perform a
more detailed analysis and find that the time taken to lock a young
star to its disc should be at most 4$\times 10^5$ yr, much shorter
than that claimed by \cite{hartmann02}, and less than the $\sim 1$ Myr
age of the ONC.  It is therefore an open question whether theoretical
predictions prefer scenario one or two.

A challenge to the explanation of rotation rates of stars in Orion
through a disc locking paradigm is apparently posed by the period
distribution in the Orion flanking fields. Although the flanking
fields also show signs of bi-modality (the Hartigan dip test gives
only a 10\% chance of unimodality), the distribution of periods
themselves are noticeably different. This is surprising, because there
is no evidence that the flanking fields differ in age from the ONC.
The largest difference is that the peak of slow rotators has shifted
from 8 days to 6 days. This difference in distributions is
statistically significant (a K-S test gives a 0.5\% chance of the two
being drawn from the same parent distribution). However, this apparent
problem has a natural explanation under the disc locking paradigm. The
disc fraction as measured by K-band excess in the OFF is significantly
lower at 15\% \citep{rebull01} than that reported for the ONC at 50\%
\citep{hillenbrand97}. Given the lower disc fractions, it is
reasonable to assume that the discs in the OFF are less massive, and
therefore less efficient at braking the young stars than those in the
ONC. Under scenario one, the slow rotators in Orion would have become
unlocked earlier from these low mass discs, and have hence spun up
from eight to six days. Assuming that the radius of the star changes
as $R \propto t^{-1/3}$, a star locked to its disc at a period of 8
days would have to have been released from disc locking at around 0.6
Myr, in order to have a period of 6 days at 1 Myr. Under scenario two,
the stars in the OFF will have become locked relatively late to their
less massive discs, and have not yet had time to become fully locked.
Using the data from Orion alone, it is not possible to distinguish
between the two scenarios.

One possible way to resolve this degeneracy is to look at the rotation
rates for older clusters. If discs are initially locked, and then
released after $\sim 1$Myr, then we would expect stars to spin-up, and
the period distribution should be more strongly biassed towards fast
rotators than seen in the ONC. Alternatively, if disc locking only
becomes effective after about 1 Myr, then the period distribution of
older clusters should show fewer fast rotators than seen even in the
ONC. Unfortunately, the period distributions of older clusters are
contradictory.  IC 348 shows a large number of slow rotators (around 8
days), and very few fast rotators (there are no period found below 2
days). This dearth of fast rotators was first commented on by
\cite{herbst00}, but their result was not significant, due to small
number statistics, and a possible lack of sensitivity to the shortest
periods. To assess the significance of the lack of fast rotators in
this study, we selected twenty high-mass stars at random from the ONC
dataset. This produced a sub-sample of the ONC data which was the same
size as our high-mass sample in IC 348. In 98\% of tests this
sub-sample had one or more stars with periods below two days. We
therefore conclude that the lack of fast rotators in IC 348 is
significant. Taken in isolation, this result would suggest that the
ONC, OFF and IC 348 data could be explained in a cohesive picture of
disc-locking, in which disc locking took $\sim 1$ Myr to become
effective. At the age of IC 348 (approx 3 Myr) disc locking has
further removed fast rotators, and increased the percentage of stars
rotating near 8 days. However, there are two problems with this
picture. The first problem is posed by the lack of any correlation
between IR excess and rotation rate in IC 348, and the uncertainty as
to the true cause of the correlation found in the ONC. A lack of a
correlation between signatures of disc presence and rotation rate is a
major challenge to disc-locking theory. The second problem is posed by
the period distribution of NGC 2264. At a similar age and disc
fraction to IC 348, we might expect NGC 2264 to show similar rotation
properties.  Instead, the period distribution in NGC 2264 is more
strongly skewed towards fast rotators than the ONC!  K-S tests reveal
that whilst the NGC 2264 distribution is not statistically different
from the OFF, it is different from the ONC and IC 348 (at 96\% and
99.6\% confidence levels respectively). This result is in direct
opposition with the result for IC 348, and causes serious difficulties
for disc locking. Given that NGC 2264 and IC 348 show very similar
ages distributions and disc fractions, disc locking theory provides no
explanation for the marked difference in period distributions.  We
must therefore also conclude that if disc locking does regulate
angular momentum in young stars, differences between cluster
environments mask many of the observational effects of that regulation.

\subsection{Stars with masses less than 0.25 M$_{\odot}$}
\label{subsec:lomass}

In both IC 348 and the ONC, the period distribution of the low-mass
stars is markedly different to that of the high mass stars. Why this
should be the case is uncertain. Interpreted in the context of disc
locking, either the low-mass stars are not locked to their discs, are
locked to their discs for a short amount of time, or are locked to
their discs at shorter locking periods than the high mass stars.
Examination of the period distributions for IC 348 and the ONC seems
to rule out this latter possibility. If we assume that IC 348 evolved
from an ONC-like period distribution then it is clear that the high
mass stars have spun down, whilst the low-mass stars have spun up.
This is not consistent with the low-mass stars having their angular
momenta regulated in the same way as the high-mass stars. Short disc
lifetimes, or a mechanism to reduce the effectiveness of disc locking,
are therefore necessary to explain the period distributions of the low
mass stars.  This implies either a difference in magnetic field
properties, or in disc properties between the high and low mass
stars. Currently, little is known about the magnetic fields in young
stars, or how they behave with stellar mass.  However, disc truncation
by stellar encounters suggests a difference in disc properties between
low and high mass stars. If low-mass stars are the results of frequent
stellar encounters, they will possess relatively low mass discs
\citep{bate03}. These discs will be inefficient at regulating stellar
angular momentum. We might therefore expect the low-mass stars angular
momentum evolution to be characterised by steady spin-up. This is the
case in IC 348 - although the high mass stars show a dearth of fast
rotators compared to the ONC, the low mass stars in IC 348 are
rotating very rapidly indeed, with the vast majority of stars having
periods around 1--2 days.  The obvious conclusion is that the discs
around low-mass stars do not regulate the stellar angular momentum in
the same way as discs around high mass stars do. An excellent test of
this suggestion will be the low-mass period distribution for NGC 2264.
Even though the high mass period distribution is very different to IC
348 the above hypothesis suggests that the low-mass distribution will
be quite similar.

\subsection{Disc locking and infrared excess}
\label{subsec:irxs}

A major challenge to any interpretation of period distributions in
terms of disc locking is presented by the lack of any reliable
correlation between disc indicators such as infrared excess and
rotation rate in any of the clusters presented here.  In the absence
of any such correlation it is tempting to say that disc locking cannot
be resonsible for regulating the angular momentum of young stars. This
is not necessarily the case, however, and two alternative
possibilities are outlined below.
\begin{itemize}
\item{The intrinsic spread in indicators might mask any correlation between
    rotation rate and rotational period. Inclination effects and the
    effects of inner disc holes can produce a large spread in the size
    of any infrared excess \citep{h98}.}
\item{The disc indicators could be {\em too sensitive}. If high mass discs
    are needed to effectively regulate a star's rotation, then the
    presence of an K-band or L-band excess does not necessarily imply
    a disc of sufficient mass.  Discs with masses as small as
    10$^{-5}M_{\odot}$ can produce a detectable K-band excess
    \citep{wood02}.}
\end{itemize}

\section{Conclusions}
\label{sec:conclusions}

We have presented a photometric study of rotation rates in the young
cluster IC 348. The depth and temporal density of our observations has
allowed us to (i) more than double the known number of periodic stars
associated with the cluster, (ii) discover periods in stars with
masses less than 0.25$M_{\odot}$ and (iii) discover periods in
classical T-Tauri stars for the first time in this cluster. The period
distribution of stars with masses greater than 0.25$M_{\odot}$ shows
the same bi-modal behaviour as seen in the ONC, although there is a
statistically significant lack of rapidly rotating stars with respect
to Orion.

We have also presented a discussion of our data in the context of previously
published period distributions for the ONC, Orion flanking fields and
NGC 2264. We find two significant problems with the disc locking paradigm.
\begin{itemize}
  
\item{We find that the previously claimed correlation between infrared
    excess and rotation period in the ONC might in fact result from a
    correlation between infrared excess and {\em mass}.}

\item{We also find a marked difference in period distributions between
    NGC 2264 and IC 348, which presents a serious challenge to the
    disc locking paradigm, given the similarity in ages and disc
    fractions between the two clusters.}

\end{itemize}

\section*{\sc Acknowledgements}
SPL is supported by PPARC. The authors acknowledge the data analysis
facilities at Exeter provided by the Starlink Project which is run by
CCLRC on behalf of PPARC. We thank W. Herbst for providing the data in
\cite{herbst04} prior to publication, and J. Pringle for useful
discussions.

\bibliographystyle{mn2e}
\bibliography{abbrev,refs}

\end{document}